\documentclass[%
 aps, pra, twocolumn,
 showpacs,
 longbibliography,
 superscriptaddress,
 amsmath,amssymb,
 floatfix
]{revtex4-1}

\usepackage{graphicx}
\usepackage{dcolumn}
\usepackage{bm}
\usepackage[colorlinks=true, allcolors=blue]{hyperref}
\usepackage{utils}
\usepackage{float}

\usepackage{silence}
\begin{document}

\title{Numerical Investigations of Electron Dynamics in a Linear Paul Trap}
\author{Andris Huang}
 \email{andrewhz@berkeley.edu}
 \affiliation{Department of Physics, University of California, Berkeley, Berkeley, CA 94720, USA}
 \affiliation{Challenge Institute for Quantum Computation, University of California, Berkeley, Berkeley, CA 94720, USA}
 
\author{Edith Hausten}
 \email{edith.hausten@stud.uni-greifswald.de}
 \affiliation{Institute of Physics, University of Greifswald, Greifswald 17498, Germany}
 
\author{Qian Yu}
 \affiliation{Department of Physics, University of California, Berkeley, Berkeley, CA 94720, USA}
 \affiliation{Challenge Institute for Quantum Computation, University of California, Berkeley, Berkeley, CA 94720, USA}
 
\author{Kento Taniguchi}
\affiliation{Komaba Institute for Science (KIS), The University of Tokyo, Tokyo 153-8902, Japan}

\author{Neha Yadav}
 \affiliation{Department of Physics, University of California, Berkeley, Berkeley, CA 94720, USA}
 \affiliation{Challenge Institute for Quantum Computation, University of California, Berkeley, Berkeley, CA 94720, USA}
 
\author{Isabel Sacksteder}
 \affiliation{Department of Physics, University of California, Berkeley, Berkeley, CA 94720, USA}
 \affiliation{Challenge Institute for Quantum Computation, University of California, Berkeley, Berkeley, CA 94720, USA}
 
\author{Atsushi Noguchi}
 \affiliation{Komaba Institute for Science (KIS), The University of Tokyo, Tokyo 153-8902, Japan}
 \affiliation{RIKEN Center for Quantum Computing (RQC), Wako, Saitama 351-0198, Japan}
 \affiliation{Inamori Research Institute for Science (InaRIS), Kyoto-shi, Kyoto 600-8411, Japan}
 
\author{Ralf Schneider}
 \affiliation{Institute of Physics, University of Greifswald, Greifswald 17498, Germany}
 
\author{Hartmut H\"{a}ffner}
 \email{hhaeffner@berkeley.edu}
 \affiliation{Department of Physics, University of California, Berkeley, Berkeley, CA 94720, USA}
 \affiliation{Challenge Institute for Quantum Computation, University of California, Berkeley, Berkeley, CA 94720, USA}
 \affiliation{Computational Research Division, Lawrence Berkeley National Laboratory, Berkeley, CA 94720, USA}
\date{\today}

\begin{abstract}
Trapped electrons have emerged as an interesting platform for quantum information processing due to their light mass, two-level spin states, and potential for fully electronic manipulation. Previous experiments have demonstrated electron trapping in Penning traps, Paul traps, on solid neon, and superfluid films. In this work, we consider electrons confined in Paul traps, with their spin states as the qubits. For this approach, if the two electrons are trapped in the same potential well, they must form Wigner crystals and remain stable under a static magnetic field to enable two-qubit gates, achievable only within certain trapping parameters. To identify feasible operating conditions, we performed numerical simulations of electron dynamics in linear Paul traps, finding the threshold temperatures required to form two-electron Wigner crystals and studying how the thresholds scale with trap frequencies. In addition, we numerically verified the cooling methods required to reach the crystallization thresholds. Lastly, we examined the stability of electrons under various magnetic field strengths and identified stable regions of trap operation.
\end{abstract}

\maketitle

\section{Introduction} \label{sec:intro}
Among all ongoing efforts to implement physical quantum processors, architectures based on trapped ions in Paul traps have exhibited advantages in their long coherence time, flexible connectivity, and low error rate per gate \cite{moses2023racetrack, delaney2024gridtrap}. These natural advantages of ion traps motivated the idea of using trapped electrons for quantum information processing, where the electrons are confined in Paul traps and their spins are used as the two-level qubits. Several properties of electrons may solve the challenges faced by scaling ion-trap processors. The lighter electron mass, being four orders of magnitude smaller compared to ions, leads to an increase of the motional frequencies in the trapping potential, thus allowing for faster multi-qubit operations and transport. Electrons are natural two-level systems, preventing information leakage from the computational subspace and potentially leading to higher-fidelity operations and simpler quantum error correction schemes. In addition, the electron spins can be controlled purely electronically using microwave pulses and static or oscillating magnetic field gradients, which reduces the optical complexity required for trapped-ion systems \cite{YuFeasibility}. 

Previous efforts have experimentally demonstrated trapping electrons in room-temperature Paul traps \cite{matthiesen2021trapping, taniguchi2025imageCurrent}, and the next step towards realizing a trapped-electron quantum processor is demonstrating the cooling of single electrons, the ability to detect spin states, and coherent manipulations of single and two electrons. We consider using the M\o lmer-S\o rensen gate and its controlled phase gate variations to conduct two-qubit operations on trapped electrons. Specifically, we apply an AC magnetic field gradient at a frequency slightly detuned from the axial center-of-mass (COM) mode frequency. As a result, a spin-dependent force acts on the electrons and a COM motion will be excited if and only if the two electrons have the same spin state \cite{YuFeasibility}. In this scheme, the electrons need to satisfy two major requirements: (1) form a Wigner crystal in the same quadratic potential well and maintain the crystal state for a duration much longer than the gate time and (2) have non-degenerate spin states. To achieve the first requirement, the two electrons need to be cooled down below a certain temperature threshold, and it is critical to find the temperature thresholds for different trap parameters. To achieve the second requirement, a static magnetic field needs to be applied to split the degeneracy, while the electron trajectories need to remain stable under the magnetic field.

When the trapped electrons have high kinetic energy, they are in a cloud state where the displacement of individual electrons from their equilibrium position can be greater than the distance between them. As the electrons cool down into a sufficiently low-energy regime, they form Wigner crystal states where the electrons are localized near their equilibrium positions and follow well-defined normal modes, leading to a low electron-chain reordering rate. As a result, analogous to the Coulomb crystals studied in ion-trap systems \cite{VanMourik2022Rf-inducedIons}, the Wigner crystal state is required for trapped electrons to conduct two-qubit operations if the two electrons are in the same trapping site.

Previous works have experimentally demonstrated the formation of large ion crystals in Paul traps \cite{drewsen1998ionCrystal, d2021ionCrystal} as well as for electrons on superfluid Helium \cite{rousseau2009Wigner, schusterWigner2019}. But electron Wigner crystals have not yet been observed in Paul traps because of a lack of sufficient cooling. Since electrons do not have internal structures, cooling them to a low-energy state requires a low-temperature environment, where the temperature must be carefully chosen to enable the formation of Wigner crystals. We expect that an energy threshold exists, below which the electrons remain crystallized for a sufficiently long duration, but this threshold is difficult to find analytically due to the presence of micromotion and the nonlinear Coulomb interaction. Past theoretical work has derived the thresholds for multi-electron Wigner crystals \cite{Vu2020} and multi-ion Coulomb crystals \cite{Hudson2013ionCrystal}, but the dynamics for few-electron crystals have not yet been analyzed, particularly with the micromotion adding complexity to the system. At large motional amplitudes, the micromotion in radial directions can cause energy transfer between the Coulomb potential, radio-frequency (RF) drive, and electrons' secular motion, hence causing heating over time. This process is called RF heating and is known from ion trap experiments \cite{VanMourik2022Rf-inducedIons}. Therefore, a time-domain numerical simulation is beneficial for studying the transition between the crystal and cloud states and finding an energy threshold.

To cool the electrons below the required threshold, a quantum charge-coupled device (QCCD) architecture is proposed here to cool two single electrons in separate potential wells and merge them into the same potential well for two-qubit operations. The cooling of electrons is typically done by coupling the electron's motion to a low-temperature tank circuit, but the tank circuit only couples to the center-of-mass (COM) mode of an electron chain and cannot effectively reduce the energy of the relative motions. As a result, it is very challenging to start with two electrons in the same well and directly cool all the modes below crystallization thresholds, and thus we chose the QCCD architecture. Since this architecture has not been experimentally realized for trapped electrons, we conducted a numerical simulation to verify that this method can satisfy the requirements for Wigner crystal formation.

In addition, for electron spin qubits, the qubit frequency is determined by the external magnetic field strength, due to the lack of atomic structures, and the qubit frequency degenerates with the cyclotron frequency. Since the mass of electrons is small compared to that of ions, at similar qubit frequencies, electrons' cyclotron frequencies are much larger than those of ions. Thus, the stability conditions for ions may not apply to electrons if a similar qubit frequency is used. Moreover, at large qubit frequencies, the electrons' cyclotron frequencies can be close to the oscillation frequencies, causing the stability conditions to potentially change from the conventional studies of ions in Paul traps, based on past works in combined Paul traps \cite{li1992combinedPaul,foot2018combinedPaul,huang1997combinedPaul}. Therefore, the stability of trapped electrons under a static magnetic field may not be guaranteed, and numerical simulations can be helpful in determining the stability region.

This paper focuses on the numerical investigations that find and validate the requirements above and the methods proposed to achieve these requirements. Sec.~\ref{sec:theory} outlines the numerical setup in this work. In Sec.~\ref{sec:threshold temperature}, the threshold temperature is first found for a fixed set of trap parameters, then the scaling of the threshold is found under variations of the trap parameters. Sec.~\ref{sec:cooling} proposes a cooling strategy for achieving the threshold and numerically validates its effects. Sec.~\ref{sec:B field} presents the dynamics of single electrons under the influence of a static magnetic field and investigates the stability of electron trapping at different field strengths. The final section~\ref{sec:conclusion} contains conclusions, perspectives, and an outlook for future work. Additional details are presented in the appendices.

\section{Device and numerical setup} 
\label{sec:theory}

\subsection{Device background}
In this work, we consider a setup outlined in our previous feasibility study of using trapped electrons as qubits \cite{YuFeasibility}. In this setup, we discuss using a linear Paul trap residing in a cryogenic environment at 0.4\,K. The trapping of charged particles in three spatial directions $x, ~y, ~z$ can be realized by applying a DC harmonic potential 
\begin{eqnarray} \label{eq:DC_potential}
    \Phi_{\rm{dc}} = \kappa U_{\rm{dc}} \left(\frac{2z^2-x^2-y^2}{2z^2_0}\right),
\end{eqnarray}
and a time-varying RF potential
\begin{eqnarray} \label{eq:RF_potential}
    \Phi_{\rm{rf}}(t) = V_0 \cos (\Omega_{\rm{rf}}t+\phi) \left(\frac{x^2-y^2}{2r^2_0}\right),
\end{eqnarray}
where $\kappa, ~z_0, ~r_0$ are the geometry-dependent trap parameters, $U_{\rm{dc}}$ and $V_0$ are the voltage amplitudes of the DC and RF potential, and $\phi$ is the initial phase of the RF drive, with a drive frequency of $\wrf$. 
By comparing Eq.~\ref{eq:DC_potential} and \ref{eq:RF_potential} to the lowest order pseudo potential approximation, we express the secular frequencies of the radial ($x,y$) and axial ($z$) directions for a particle with charge $q$ and mass $m$ as:
\begin{gather}\label{eq:trap_frequencies}
    \omega_{x,y} = \frac{|qV_0|}{\sqrt{2}m\Omega_{\rm{rf}}r^2_0}, ~\omega_{z} = \sqrt{\frac{2q\kappa U_{\rm{dc}}}{mz_0^2}},
\end{gather}
which enables us to express the stability parameters in terms of trap frequencies as 
\begin{align}
    &|q_{x,y}| = \frac{2\sqrt{2}\omega_r}{\wrf}, ~|a_{x,y}| = \frac{1}{2}|a_z| = \frac{2\omega_z^2}{\wrf^2},
\end{align}
where $\omega_r = \omega_x = \omega_y$ represents the degenerate radial secular frequency and $\omega_z$ the axial frequency. Throughout this paper, we mainly focus on the frequency configurations analyzed in the prototype design in \cite{YuFeasibility}, where $\wrf/2\pi = 10.6$\,GHz, $\omega_r/2\pi = 2$\,GHz, and $\omega_z/2\pi = 300$\,MHz.

\subsection{Numerical integration algorithms}
In this work, the electron trajectories $\mbf{r}(t)$ are obtained numerically by solving the equations of motion of the form 
\begin{eqnarray}
    \Ddot{\mbf{r}} = \frac{1}{m}\sum_i\mbf{F}_i(\mbf{r}, \Dot{\mbf{r}}, t).
\end{eqnarray}
The force components $\mbf{F}_i$ involved vary in each section depending on the specific application. The common force present throughout all scenarios is the trapping force originating from the trap potential described in the previous section in Eq.~\ref{eq:DC_potential} and Eq.~\ref{eq:RF_potential}. The trapping force can thus be found through
\begin{eqnarray}
    \mbf{F}_{\rm{trap}}(\rb, t) = -q\Del \left[\Phi_{\rm{dc}}(\rb)+\Phi_{\rm{rf}}(\rb, t)\right].
\end{eqnarray}
In Sec.~\ref{sec:threshold temperature}, our major interest is to study the interactions between two electrons in the same trap, therefore we need to consider the Coulomb interaction $\mbf{F}_{\rm{Coul}}(\rb_1, \rb_2)$. In Sec.~\ref{sec:cooling}, the cooling and noise forces $\F_{\rm{damp}}(\rdot)$ and $\F_{\rm{noise}}(t)$ need to be considered, as well as additional forces arising from time-dependent potentials to conduct parametric coupling and electron transportation. In Sec.~\ref{sec:B field}, the Lorentz force $\F_{\rm{B}}(\rdot)$ needs to be considered since we are studying charged particle dynamics in magnetic fields. Note that among all the forces involved, the damping and Lorentz forces are dependent on the electron's velocity, while the other forces depend on the electron's position and time.

We have used three different numerical algorithms within this work: the Velocity Verlet integration (VV) \cite{hairer2003verlet}, the third-order Runge-Kutta (RK3) integration \cite{bogacki1989RK23}, and the Alternating Frequency-Time domain Harmonic Balance (AFT-HB) method \cite{Zipu2023, Nicks2024}. The VV algorithm is a fast second-order algorithm commonly used in simulating particle dynamics, but it cannot be used in cases where the forces are velocity-dependent \cite{chambliss2020VelocityVerlet}. Thus, in Secs.~\ref{sec:cooling} and \ref{sec:B field} where the damping or Lorentz forces are involved, we used the RK3 algorithm instead. Finally, the AFT-HB method is a semi-analytical approach that conducts simulations alternatively in frequency and time domain, which enables us to analyze the transition between stable and unstable motions much more efficiently than the traditional integration algorithms \cite{Zipu2023, Nicks2024, kentoFloquetEngineering}. Since usage of the AFT-HB method is relatively new for Paul trap studies, this work also extends the range of applicability of the method and demonstrates the potential for faster analysis of particle dynamics in Paul traps.

In all simulations using the VV and RK3 algorithm, we chose a fixed time step of $dt=10^{-13}$\,s, which is much shorter than the period of all oscillation modes in the system, including micromotion. This choice of the time step ensures that all the fast-moving dynamics can be captured with high accuracy.

\section{Threshold Temperature for Two-Electron Wigner Crystals}
\label{sec:threshold temperature}

\subsection{Full dynamics simulation for threshold finding}
Numerically, we simulated the trajectories of electrons assigned with different initial kinetic energies at their equilibrium position. We then record the Wigner crystal lifetime, defined as the time before the first reordering event occurs under a certain initial condition. Sample trajectories for electrons in crystal and cloud states are shown in Appendix \ref{sec:sample_traj_crystal}. The threshold kinetic energies can therefore be found by monitoring the initial kinetic energy below which the two electrons consistently do not reorder for a desired duration. Connecting to our motivation of having high-fidelity two-qubit operations, we want the electrons to remain crystallized for a duration much longer than the gate time, which is expected to be around 2\,\textmu s \cite{YuFeasibility}. Therefore, the simulations in this section were all conducted for a time duration of 1\,ms.

In our setup, since $\omega_r \gg \omega_z$, the confinement in the radial direction is much stronger than that in the axial direction. Hence, for the few-electron case, the electrons will form a linear chain along the axial direction. From the harmonic potential in $z$ direction as in Eq.~\ref{eq:DC_potential}, we can find the equilibrium distance between two electrons as
\begin{equation}
    l = \left(\frac{q^2}{2\pi\epsilon_0 m\omega_z^2}\right)^{1/3}
\end{equation}
The electrons' initial positions are thus along the $z$ direction separated by $l$. 

Here, we consider the full 3D model where two electrons oscillate in both radial and axial directions. The two-electron dynamics can be decomposed into a COM and relative (stretch) mode for the axial and radial directions each, hence yielding 6 total normal modes. However, only 4 modes have different frequencies as the two radial modes $\omega_x$ and $\omega_y$ are degenerate. We expect that the relative motion contributes the most to the crystal-cloud transition as the COM motion does not change the spacing between the electrons. Therefore, we define the threshold energy as the maximum initial kinetic energy that the relative motion of electrons along the axial or radial direction can have for a Wigner crystal lifetime longer than 1\,ms. The axial and radial threshold energies are determined separately by varying the initial kinetic energy in the respective direction while keeping all other modes cold at the energy $\Ekin = \kB T/2$, with $T=0.4$\,K being the proposed temperature for the cryogenic environment. 

\begin{figure*}
    \centering
    \includegraphics[width=0.96\textwidth]{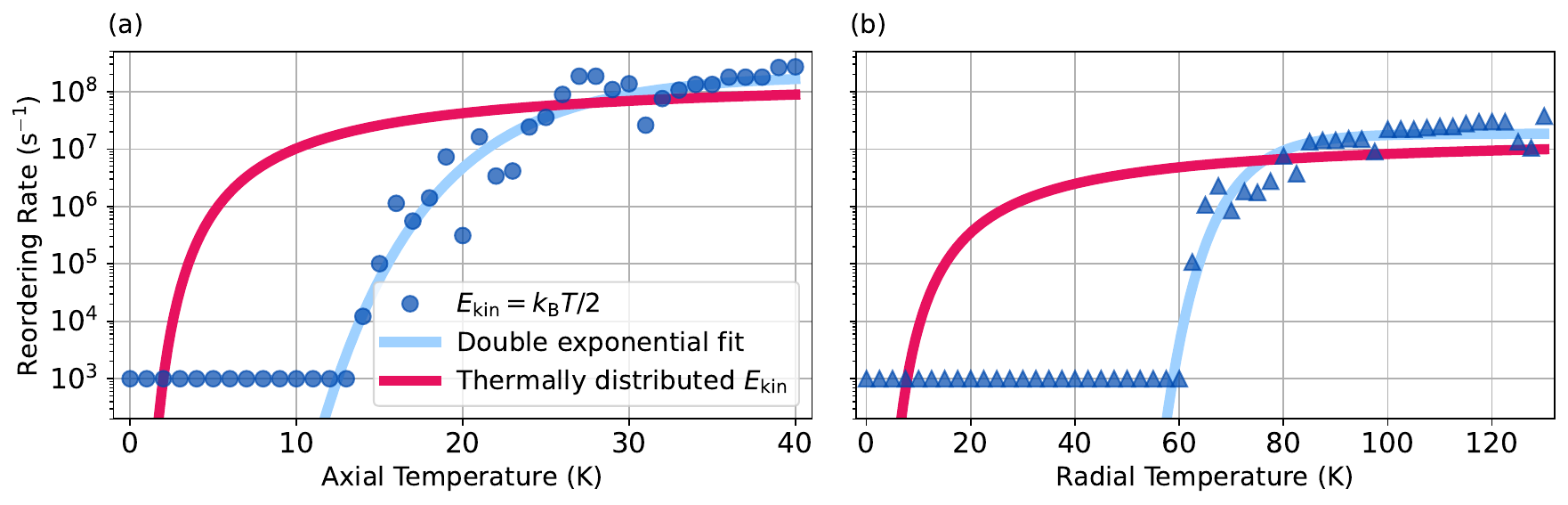}
    \caption{Two-electron chain reordering rate with respect to different initial (a) axial and (b) radial energies. Each blue data point represents a simulation run with a distinct initial energy. A double exponential fit, shown in light blue, is used to extrapolate the reordering rate $f_{\rm{reorder}}$ beyond the simulation limits. The mean reordering rate $\Bar{f}_{\rm{reorder}}$ given a thermal Boltzmann distribution of electron energies is then computed based on the extrapolated reordering rates and is shown in red. The horizontal axes are labeled in units of Kelvin for simpler conversions between the energies and the temperature of the environment. The simulation is conducted with the drive frequency at $\wrf/2\pi$ = 10.6\,GHz, radial frequency at $\omega_r/2\pi=2$\,GHz, and axial frequency at $\omega_z/2\pi=300$\,MHz.}
    \label{fig:threshold_temperature}
\end{figure*}

The obtained result is shown in Fig.~\ref{fig:threshold_temperature}, where the electron reordering rate $f_{\rm{reorder}}$ is plotted with respect to the temperature $T$. The reordering rate is defined as the inverse of the Wigner crystal lifetime. Each blue data point represents a single simulation run with a distinct initial kinetic energy of $\Ekin = k_BT/2$. The simulated reordering rate is lower bounded by $10^3$\,\rate, due to the maximum simulation time of 1\,ms. A double exponential fit of the form $f_{\rm{reorder}}(E)=\exp\left\{A\left[1-\exp\left((E_0-E)/\tau\right)\right]+f_0\right\}$ is performed to extrapolate the reordering rate beyond the 1\,ms limit, with fitting parameters $A, ~E_0, ~\tau, ~f_0$. We also show the mean reordering rate $\Bar{f}_{\rm{reorder}}$ assuming that the thermal energies of trapped electrons follow a Boltzmann distribution $P(E, T)\propto \exp\left\{-E/(\kB T)\right\}$. We calculated the mean rate by weighting the simulated reordering rate with the Boltzmann distribution, with the form $\Bar{f}_{\rm{reorder}}(T)=\int_0^\infty P(E, T)f_{\rm{reorder}}(E)~dE$. The resulting curve is plotted in red, and we found that for a mean reordering rate below $10^3$\,\rate, the axial and radial temperatures need to be below 1.99\,K and 7.82\,K, respectively. The rate threshold of $10^3$ was chosen as a compromise between long desired Wigner crystal lifetimes of more than 1\,ms and the computational cost to simulate for such long times. We note that assuming anticipated gate times of 2\,\textmu s \cite{YuFeasibility}, crystal lifetimes of 20\,ms are required to achieve error rates of ${10^{-4}}$.

\subsection{Threshold scaling with trap parameters}\label{sec:threshold_finding}
In the classical limit, the Wigner crystal formation requires the average kinetic energy $\expval{E_{\rm{kin}}}$ to be smaller than the average Coulomb potential energy $\expval{U_{\rm{Coul}}}$ \cite{Vu2020}, where
\begin{eqnarray}\label{eq:coulomb_potential}
    U_{\rm{Coul}}(\mbf{r}_1,\mbf{r}_2)=\frac{1}{4\pi\epsilon_0}\frac{q^2}{||\mbf{r}_1 -\mbf{r}_2||},
\end{eqnarray}
is the potential Coulomb energy for a two-electron system. Evaluating the equation at the electrons' equilibrium positions, we can find an estimate of the energy threshold for crystal formation as 
\begin{eqnarray}
    \expval{E_{\rm{kin}}}\leq \frac{1}{2}\left(\frac{q^2}{2\pi\epsilon_0}\right)^{2/3}\left(m\omega_z^2\right)^{1/3}.
\end{eqnarray}
\begin{figure}
    \includegraphics[width=0.47\textwidth, left]{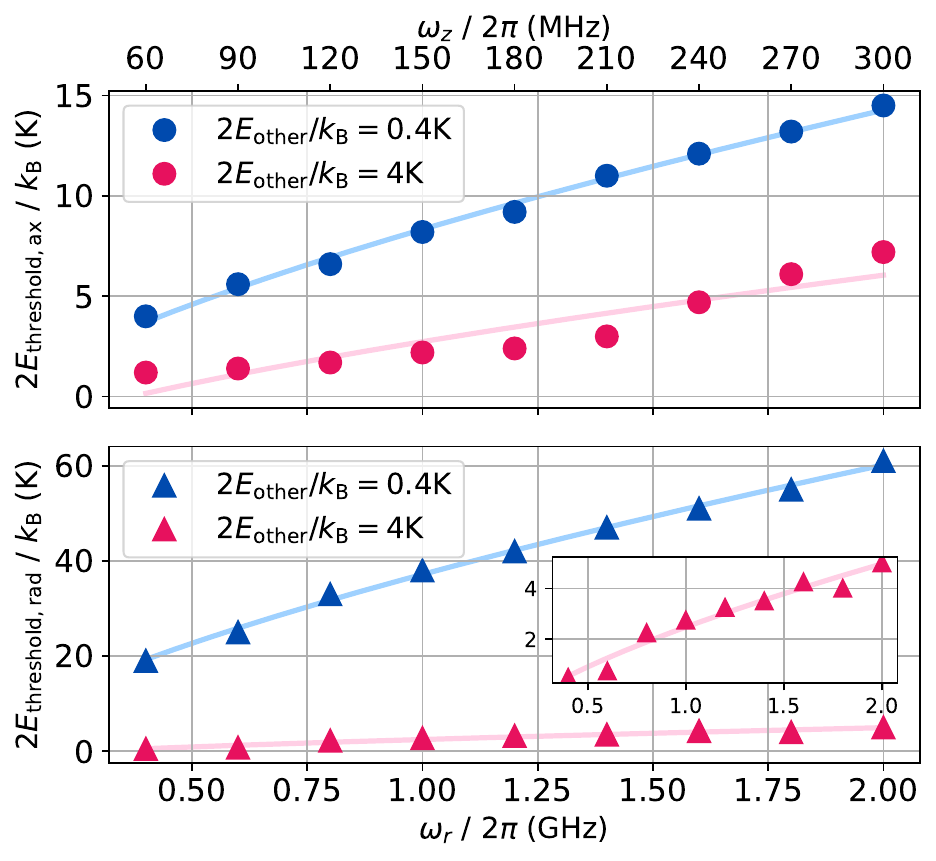}
    \caption{Axial (top) and radial (bottom) threshold energies required for two electrons to stay crystallized for 100\,\textmu s. Each
    data point represents a simulation scan as in Fig.~\ref{fig:threshold_temperature} with a distinct motional frequency. The solid curves correspond to a fitting of $2E_{\text{threshold},i}/k_{\rm{B}} = A\omega_i^{2/3} - E_{0,i}$. In each scan, the ratio between the RF drive frequency and the motional frequencies are kept constant, with stability parameters of $q=0.53$ and $a_z=0.003$. When the axial (radial) mode energy is varied, all other modes are kept constant at an equivalent temperature of 0.4\,K (blue) and 4\,K (red). The inset shows a zoomed-in version of the radial threshold energies when the other modes are at 4\,K, for a better contrast.}
    \label{fig:frequency_scaling}
\end{figure}
We expect that the COM modes are decoupled from the stretch modes in the temperature range of interest, but the radial and axial stretch modes are coupled to each other due to the nonlinear Coulomb interaction. Therefore, we expect that the radial and axial stretch mode excitations will simultaneously determine whether the electrons form a crystal and a lower limit for the respective threshold energies scale with the secular frequency as
\begin{eqnarray}\label{eq:threshold_scaling_frequency}
    E_{\text{threshold}, i} \leq A\omega_i^{2/3} - E_{0,i},
\end{eqnarray}
where $i\in {x,y,z}$ is the direction, $A$ is a proportionality parameter absorbing all fundamental constants, and $E_{0,i}$ represents the energy contributions of all other modes. We also expect that if the relative modes are sufficiently cooled, they can be decoupled from each other, and the crystals can live longer.

\begin{figure}
    \includegraphics[width=0.47\textwidth, left]{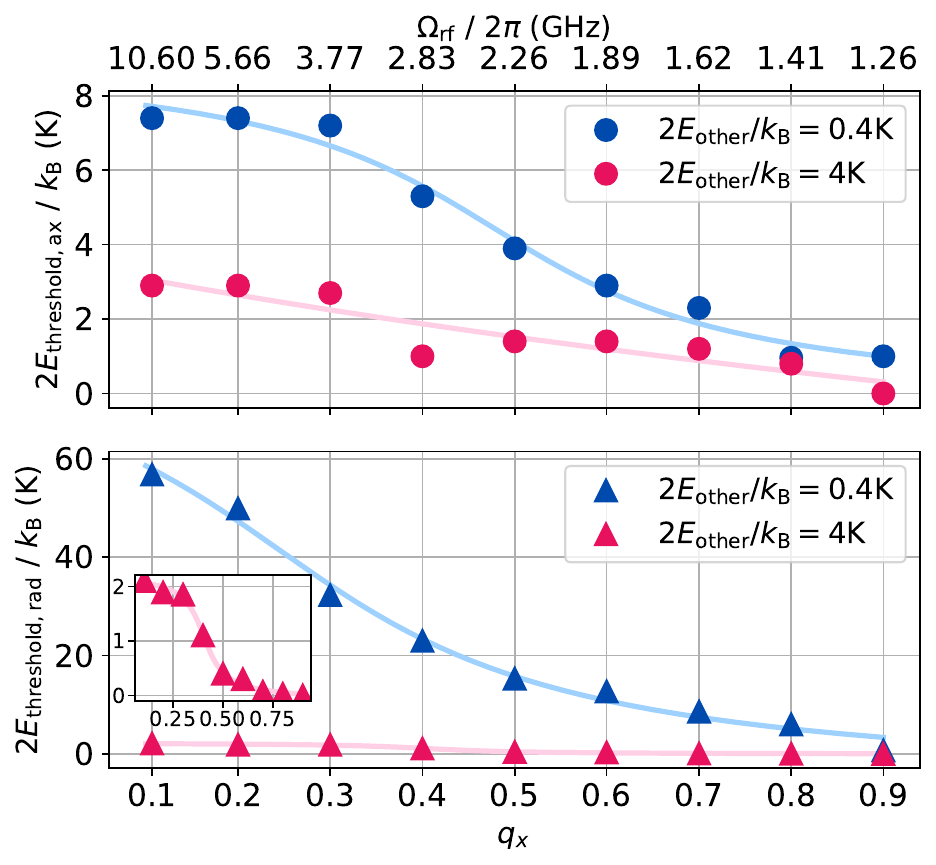}
    \caption{Axial (top) and radial (bottom) threshold energies required for two electrons to stay crystallized for 100\,\textmu s. In each run, the axial and radial frequencies are kept constant, and the RF drive frequency is varied, from which the values of $q_x$ are computed. When the axial (radial) stretch mode energy is varied, all other modes are kept constant at 0.4\,K (blue) and 4\,K (red). The inset shows a zoomed-in plot of the radial thresholds when all other modes are kept at 4\,K for a better contrast. The solid curves correspond to the fitting of the data with an inverse tangent function.}
    \label{fig:qx_scaling}
\end{figure}

We investigated the scaling of threshold energy with respect to trap frequencies by repeating 100\,\textmu s simulations for different $\omega_r, ~\omega_z,$ and $\Omega_{\rm{rf}}$. To ensure that the stability of trapping does not change for different configurations, we first varied the three frequencies simultaneously while keeping their ratio constant, i.e. maintaining $q_x = 0.53$ and $|a_z| = 0.003$. While scanning the threshold energy in one direction, we kept all other modes cold at a fixed energy. As shown in Fig.~\ref{fig:frequency_scaling}, the threshold energy increases with larger motional frequencies, and the scaling follows the $T\propto \omega^{2/3}$ relation derived in Eq.~\ref{eq:threshold_scaling_frequency}. To investigate the impact of energy contributions from other modes, we performed two sets of simulations: one with all other modes held at 0.4\,K, and the other at 4\,K. It can be seen from Fig.~\ref{fig:frequency_scaling} that the energy contribution from other modes lowers the threshold energies of the relative mode.
Since micromotion is one of the main heating sources, it is also important to study its influence on the crystal thresholds. Here, we examined the effect by varying the stability parameter in the simulations, while the motional frequencies were kept constant and the drive frequency was varied. We conducted simulations for 100\,\textmu s and recorded the threshold energy for different stability parameters $q_x$. For numerical efficiency, we set the largest drive frequency to be $2\pi\times 10.6$\,GHz and kept the time step $dt=10^{-13}$\,s to ensure the accuracy of the algorithm. To scan a large range of $q_x$ values from 0.1 to 0.9 under these restrictions, we chose a set of low motional frequencies of $\omega_r/2\pi = 400$\,MHz and $\omega_z/2\pi = 30$\,MHz. As shown in Fig.~\ref{fig:qx_scaling}, the threshold energy is higher at smaller $q_x$, where electron trapping is more stable. The effect plateaus at small $q_x$ values where the RF drive frequency is much higher than the motional frequencies, causing the micromotion to diminish. Two sets of simulations with the other modes held at 0.4\,K and 4\,K were conducted respectively, and it is worth noting that the threshold energies for the radial direction are much lower when the other modes are kept at 4\,K. In particular, when the axial mode is hot, the electrons can only tolerate low energy in the radial direction to maintain the kinetic energy below the Coulomb potential energy barrier, and the radial energies need to be near the ground state at high $q_x$ values to reduce the coupling between the radial and axial stretch modes so that the two electrons cannot exchange positions.

\section{Cooling the electrons}
\label{sec:cooling}

\subsection{Cooling dynamics}
\label{sec:cooling_dynamics}

Electrons in the trap can be cooled resistively by coupling the image current induced by the axial motion to a cryogenic tank circuit. 
The induced current will induce a voltage drop across the tank circuit, which then generates a damping force on the electron, expressed as
\begin{eqnarray}
    F_{\rm{damp}}(\dot{z})=-\left(\frac{q^2 \Re{Z}}{d_{\rm{eff}}^2}\right)\dot{z},
\end{eqnarray}
where $\dot{z}$ is the velocity in $z$ direction, $\deff$ is the effective distance assuming the trap is approximated as a parallel-plate capacitor, and $\Re{Z}=Q\sqrt{L/C}$ is the on-resonance impedance of the tank circuit \cite{YuFeasibility}. For the setup in this work, we assume $\deff=254$\,\textmu m and we consider $L=250\,\unit{nH}$, $C=1\,\unit{pF}$, resonant at $\omega_z/2\pi=300$\,MHz with $Q=1000$, leading to the on-resonance impedance of $\Re{Z} = 500\,\unit{k}\Omega$ \cite{YuFeasibility}. The corresponding cooling time constant is approximately $\tau_{\rm{c}} = 4$\,\textmu s. For this set of parameters, the bandwidth of the tank circuit is $\Delta \omega_\text{res} = \omega_z/Q = 2\pi \times 300\,\unit{kHz}$, while the axial secular frequency has a linewidth of $\Delta \omega_z = 1/\tau_{\rm{c}} = 2\pi \times 35\,\unit{kHz}$. Since $\Delta \omega_z \ll \Delta \omega_\text{res}$, we approximate the secular motion to be always on resonance with the tank circuit. Hence, the tank circuit behaves like a single resistor with $R = 500\,\unit{k}\Omega$, and the generated Johnson noise in the circuit can be modeled as a Gaussian white-noise process, which can be described as

\begin{equation}
    V_{\rm{noise}}(t) = \alpha\Gamma(t),
\end{equation}
where $\alpha$ is the strength of the noise and 
\begin{equation}
    \Gamma(t) = \lim_{\Delta t\rightarrow 0}\mathcal{N}\left(0, \frac{1}{\Delta t}\right)
\end{equation}
is a Gaussian random variable with 0 mean and variance $1/\Delta t$. In our simulation, $\Delta t $ is the integration time step $dt$. The noise strength is related to the power spectral density by $S_V = 2\alpha^2$ \cite{Gillespie1996}. The power spectral density of the Johnson noise is $S_V= 4k_{\rm{B}}T_{\rm{res}}\Re{Z}$, where $T_{\rm{res}}$ is the temperature of the circuit. Therefore, $\alpha = \sqrt{2k_{\rm{B}}T_{\rm{res}}\Re{Z}}$, and the overall voltage generated by Johnson noise is expressed by
\begin{equation}
    V_{\rm{noise}}(t) = \sqrt{2k_{\rm{B}}T_{\rm{res}}\Re{Z}}\mathcal{N}\left(0, \frac{1}{dt}\right).
\end{equation}
The force acting on the electron due to the Johnson noise can then be modeled as
\begin{eqnarray}
    F_{\rm{noise}}(t) = -\frac{q}{d_{\rm{eff}}}V_{\rm{noise}}(t).
\end{eqnarray}

We define the final temperature as the average total energy in the system after sufficient cooling periods. Assuming the Paul trap has an ideal quadratic potential, the total energy can be approximated by $\expval{E_{\rm{tot}}} = 2\expval{E_{\rm{kin}}}=m\expval{\dot{z}^2}$. At equilibrium, the energy stored in the system is determined by the Johnson noise, i.e. $k_{\rm{B}}T_{\rm{res}}$. Therefore, the final temperature of the cooled axial mode should be equal to the temperature of the tank circuit. As proof of principle, we conducted 20 simulations with the tank circuit at 0.4\,K and the electron's axial temperature initialized at the equilibrium temperature 0.4\,K. For each run, we simulated for a duration of approximately $12\,\tau_{\rm{c}}$ and calculated the final temperature by averaging the energy over the last 6 $\tau$. The average final temperature of all simulations is $0.41 \pm 0.029$\,K, which agrees with the expectations. We have also verified the effect when the electron was initialized at different temperatures.

\subsection{Cooling single electrons} \label{sec:cooling-single}
For a single electron in the trap, we study the case where the axial motional mode is considered the primary mode, and it can be cooled by attaching the cryogenic tank circuit to the axial direction of the trap. The secondary radial modes can be cooled simultaneously with the axial mode by applying an additional RF potential $\Phi_{\rm{p}}$ to parametrically couple the axial and radial modes \cite{Gorman2014}. We define the additional potential as $\Phi_{\rm{p}}(r, t) = V_1\cos(\omega_{\rm{p}}t)rz/r_0^2$, where $r=x$ or $y$ represents the degenerate radial direction to cool, $\omega_{\rm{p}} = \omega_r - \omega_z$ is the on-resonance parametric drive frequency and $V_1=A_{\rm{p}}\cdot V_0$ is the amplitude of the RF trapping potential, scaled by a factor $A_{\rm{p}}$, where $A_{\rm{p}}$ needs to be sufficiently small to avoid unwanted resonances and ensure trapping stability. Under the rotating wave approximation (RWA) and substituting $V_1 = A_{\rm{p}}V_0$, the interaction Hamiltonian has the form
\begin{align}
    H_I &\approx \frac{\hbar}{2}A_{\rm{p}}\Omega_{\rm{rf}}\sqrt{\frac{\omega_r}{2\omega_z}}(\a{r}\adag{z} + \adag{r}\a{z}) \nonumber \\
    &\equiv \hbar g_{rz}(\a{r}\adag{z} + \adag{r}\a{z}),
\end{align}
where $\a{r(z)}$ and $\adag{r(z)}$ are the annihilation and creation operators for the radial (axial) direction, and $g_{rz} = \frac{1}{2}A_{\rm{p}}\Omega_{\rm{rf}}\sqrt{{\omega_r}/{(2\omega_z)}}$ is the coupling strength between the radial and axial modes. 
Here, we chose $A_{\rm{p}} = 5.2\pwr{-6}$ to numerically demonstrate the cooling effect with $g_{rz}/2\pi = 50.3 \,\unit{kHz}$.

\begin{figure}
    \includegraphics[width=0.96\linewidth, left]{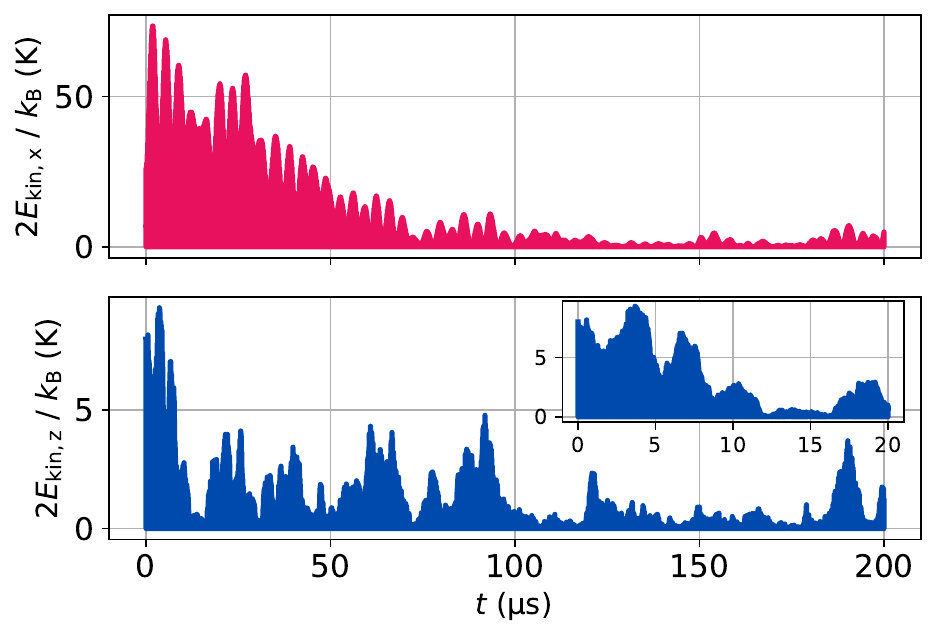}
    \caption{Simulation of single-electron cooling using the parametric drive, with the energy evolution shown for the radial direction (top) and the axial direction (bottom). The axial motion is cooled through the coupling to the 0.4\,K tank circuit, and the radial mode is parametrically coupled to the axial mode. The inset shows the zoomed-in dynamics of the first 20\,\textmu s to better present the cooling dynamics of the axial mode.}
    \label{fig:cooling-parametric_demo}
\end{figure}

We simulated the cooling of single electrons with parametric drive using a tank circuit at $0.4\,\unit{K}$ with Johnson Noise included. The energy evolution of the motion in both directions over time is shown in Fig.~\ref{fig:cooling-parametric_demo}. The results demonstrate the coupling between $x$ and $z$ directions and that the radial direction can be cooled down through the axial circuit. However, the axial cooling time is much shorter than the radial cooling time, since radial cooling is limited by the coupling strength between both modes. The inset plot within the figure shows the dynamics of the first 20\,\textmu s for the axial direction, i.e. a bit more than the cooling time constant. The equivalent temperature at thermal equilibrium is found to be 0.33\,K for the axial direction and 2.31\,K for the radial direction, for this particular simulation. The averaged values over 20 runs are $0.37\pm 0.05$\,K and $3.08\pm 0.38$\,K, respectively for axial and radial directions, which match our expectation where $T_x/T_z\approx \omega_x/\omega_z$ \cite{Gorman2014}.

We note that it is more desirable to cool the high-frequency mode, allowing one to cool the low-frequency mode below the equilibrium temperature of the tank circuit. Upon attaching the tank circuit to the radial directions, we expect that the radial modes at $\omega_r/2\pi \approx 2\,\unit{GHz}$ are cooled to 0.4\,K and coupling one of them to the axial mode at $\omega_\textrm{z}=300\,$MHz would yield a temperature of 60\,mK \cite{YuFeasibility}.

\subsection{Shuttling and merging}
Cooling a multi-electron system is difficult as the resistive cooling mechanism above is only effective for the COM motion, but not the relative motion of a given structure. This problem can be overcome by cooling single electrons in individual potential wells and transporting and merging them into a shared potential well for conducting further operations. The shuttling and merging process can be conducted by adjusting the DC potential over time. Fast shuttling with negligible heating has been demonstrated for ions in Paul traps \cite{delaney2024gridtrap}, and the process of shuttling electrons is analogous. In addition, the shuttling of electrons can be much faster than ions with increases in energy similar to those of the ions, benefiting from a smaller electron mass. Using previous results in ion trap studies, for a near-adiabatic shuttling process using a sinusoidal transport function with a timescale $\tau_{\rm{t}} \gg \omega_z^{-1}$, the average number of motional quanta transferred to the electrons during a shuttling process of 1\,\textmu s duration is around $\Delta \Bar{n}\approx 10^{-5}$ \cite{LittichThesis}.

To split and merge two-electron crystals, we consider the potential in the form
\begin{eqnarray}
    \Phi_{\rm{s}}(z, t) = \alpha(t) z^2 + \beta(t) z^4,
\end{eqnarray}
where $\alpha(t), ~\beta(t)$ are the time-dependent amplitudes that can be controlled through the DC electrodes \cite{qi2021thesis}. Since splitting and merging are time-reversal processes, we only present results for splitting processes here for simplicity. We simulated splitting dynamics using DC potential profiles as described above with different total splitting times to investigate the number of motional quanta gained in the process. The time evolution of DC potentials can be designed by first choosing an evolution profile of the electrons' equilibrium positions. Here, we consider a sinusoidal profile, which has been experimentally demonstrated for ion traps in splitting experiments \cite{kaufmann2014splitting, qi2021thesis}, where the equilibrium distance between two electrons has the form
\begin{eqnarray}
    d(t) = \frac{d_0}{2}\left[1-\cos\left(\pi\frac{t}{\tau_{\rm{s}}}\right)\right],
\end{eqnarray}
with $d_0$ the initial separation distance and $\tau_{\rm{s}}$ the total splitting time. The amplitude of the quartic confinement $\beta (t)$ can then be adiabatically ramped up from the initial value $\beta_0$ to its maximum value $\beta_{\rm{CP}}$ at the critical point (CP), using the same profile as above, then adiabatically ramped down to 0 using the time reversal of the same profile \cite{qi2021thesis}. The critical point is defined as the location where $\beta(t)$ is maximized and $\alpha = 0$, so the two electrons are only confined by the quartic potential. As a result, the frequency of the two electrons naturally drops to the lowest value at the critical point \cite{kaufmann2014splitting, qi2021thesis} and scales with $\beta_{\rm{CP}}$ as
\begin{eqnarray}
    \omega_{\rm{CP}} = \left(\frac{q}{2\pi\epsilon_0}\right)^{0.2}\left(\frac{3q}{m}\right)^{0.5}\beta_{\rm{CP}}^{0.3}.
\end{eqnarray}
We assume the lowest frequency in the process is $\omega_{\rm{CP}}/2\pi = 100$\,MHz, which requires a maximum $\beta$ value of 3\,kV/mm$^4$. This value is on the same order of magnitude as the splitting and merging experiments performed in ion traps, where a maximum $\beta$ on the order of 10\,kV/mm$^4$ has been demonstrated \cite{palmero2015splitting, wilson2014tunableMerging}. The evolution profile for $\alpha(t)$ can then be calculated using the relation between the equilibrium distance and confinement strength $\beta d^5 + 2\alpha d^3 = q/(2\pi\epsilon_0)$. The initial values of $\alpha$ and $\beta$ are determined by the initial trap frequencies, where $\beta = 0$ since the electrons are only confined by the quadratic potential and $\alpha$ can be calculated using Eq.~\ref{eq:DC_potential}, with $\omega_z/2\pi = 300$\,MHz. We chose the final separation between the two electrons to be 200\,\textmu m, and the final potential can be calculated accordingly. 

\begin{figure*}
    \centering
    \includegraphics[width=0.96\linewidth]{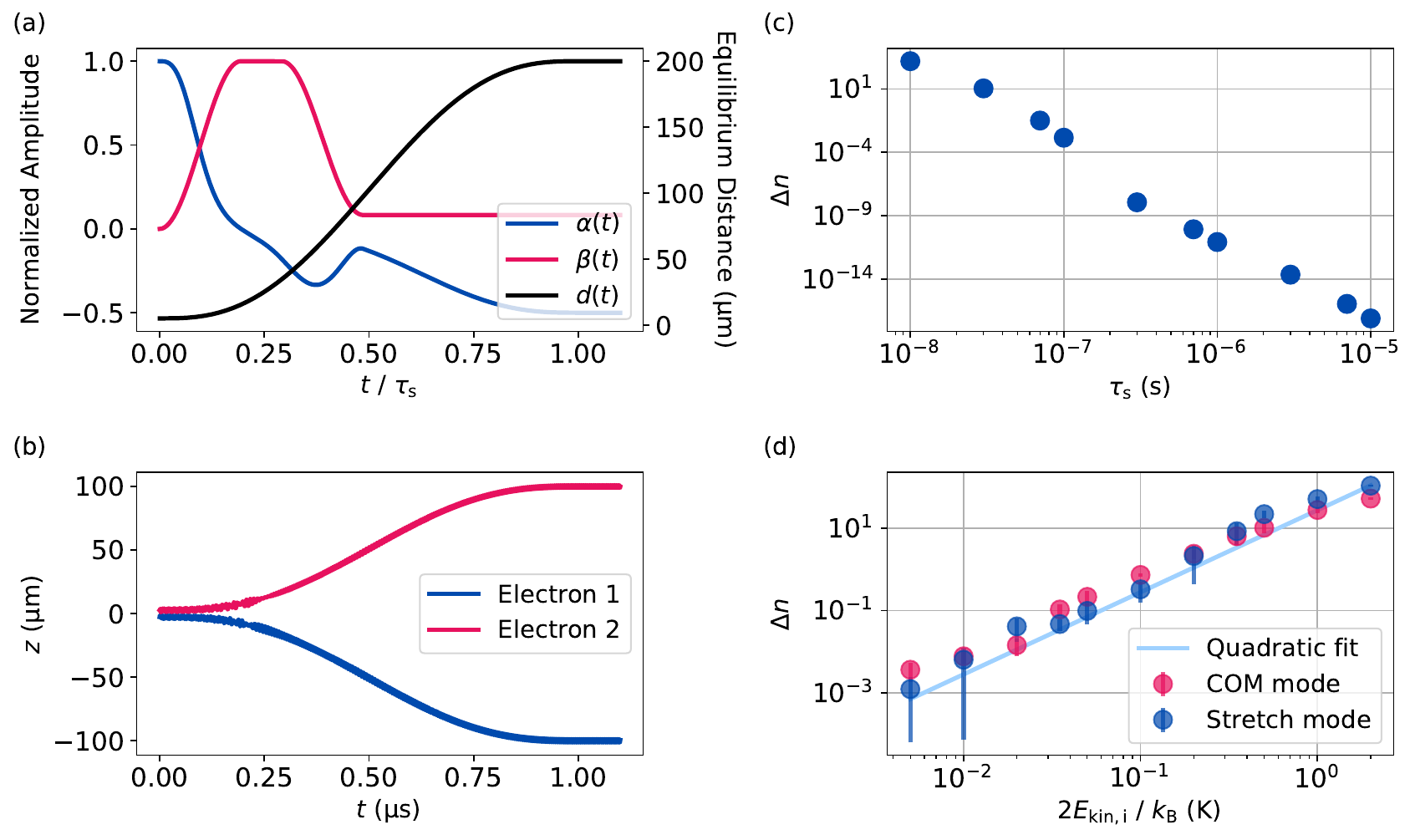}
    \caption{(a) Time evolution for DC potentials and equilibrium distance during the splitting process. (b) A sample trajectory for a splitting process with 1\,\textmu s splitting time. The two electrons' axial COM and stretch modes are both assigned an initial kinetic energy of $2\Ekin/\kB = 0.4$\,K. (c) Coherent heating induced by the splitting process for different process durations. The electrons are initialized at zero energy. (d) Heating induced during a 1\,\textmu s splitting process with different initial kinetic energy in the axial COM and stretch modes. Note that in this discussion we have neglected potential heating from finite radial mode energies. }
    \label{fig:cooling-splitting-merging}
\end{figure*}

The splitting profile used in the simulations and a sample trajectory are shown in Fig.~\ref{fig:cooling-splitting-merging}. We first performed a simulation where all axial and radial modes were initialized at zero energies. We observed that the heating in the process is negligible for a splitting time of 1\,\textmu s, with $\Delta n = 8\pwr{-12}$ quanta. We also found that, with a similar amount of heating, the time needed to split electrons is much less than ions due to the difference in mass \cite{qi2021thesis, kaufmann2014splitting}. However, one critical challenge appears as it is difficult to cool the electrons to the ground state before the splitting procedure. Therefore, we also need to investigate the heating during the process with non-zero initial energies. In Fig.~\ref{fig:cooling-splitting-merging}, for a fixed splitting time of 1\textmu s, the change in quanta is plotted with respect to the initial energy, where the axial COM and stretch modes are simultaneously assigned an initial kinetic energy while the radial modes are initialized at zero energy. We observed that the heating scales quadratically with the initial kinetic energy, and a quadratic fit is provided for the stretch mode, with the form $\Delta n =C(E_i)^2 + 8\pwr{-12}$, where $E_i$ represents the initial energy and $C$ is a fitting parameter. As a result, splitting and merging electron crystals can be a challenge in this cooling scheme, since the heating during the process becomes more prominent when the electrons are not cold at the beginning of the process. Nevertheless, if the electrons are cooled to 0.4\,K at the beginning of the merging process, the final energy is still below the crystal transition threshold, but further cooling on the axial stretch mode may be beneficial for high-fidelity gate operations.

These results also show that cooling electrons is important not only for reasons concerning gate fidelity but also to overcome heating during the transport process. Therefore, cooling methods that can efficiently cool electrons to near ground states will be meaningful in the long run. In addition, the splitting and merging dynamics can be further optimized by designing more complex electrode geometries \cite{kaufmann2014splitting, home2004splitting} or using different transport profiles \cite{qi2021thesis, palmero2015splitting}, which have also been experimentally realized in trapped ion systems.

\subsection{Cooling two electrons}
We can further cool the normal modes resistively once we merge two electrons into the same quadratic potential well. The two electrons' axial and radial COM modes can be resistively cooled using the tank circuit attached to the respective direction. However, the cooling mechanism cannot directly access the stretch modes since the resonators only couple to the COM modes of the electrons. We can overcome the problem by applying an additional RF field to couple the axial COM and stretch mode to indirectly cool the stretch mode \cite{Hou2022ParametricCoupling, Hou2023IndirectCooling}.

The coupling potential is of the form
\begin{eqnarray}\label{eq:U_z}
    \Phi_{\rm{p}}(z, t) = \left(\frac{V_2}{r_0^3}\right) \cos(\omega_{\rm{p}}t)z^3,
\end{eqnarray}
where $\omega_{\rm{p}} = \omega_{\rm{s}} - \omega_{\rm{c}}$ is the difference between the frequency of the axial COM mode and stretch mode, and $V_2$ is the amplitude of the parametric RF field. We can first rewrite the potential in the form of $\Phi_{\rm{p}}(z, t) = (V_2/r_0^3)\Phi(z)\cos(\omega_{\rm{p}}t)$, where $\Phi(z) = z^3$. Assuming the two electrons are at axial positions $\mbf{z} = (z_1,z_2)^\T\approx(z_{0,1}+\delta z_1,z_{0,2}+\delta z_2)^\T$, where $z_{0, i}$ is the equilibrium position of electron $i$ and $\delta z_i$ is the displacement from the equilibrium, the potential $\Phi (z)$ can be expanded as
\begin{eqnarray}\label{eq:cooling-expansion}
    \Phi(z_{0,i}+\delta z) \approx \Phi(z_{0,i}) + \frac{\del \Phi}{\del z}|_{z=z_{0,i}}\delta z_i \nonumber \\
    + \frac{1}{2}\cdot \frac{\del^2 \Phi}{\del z^2}|_{z=z_{0,i}}\delta z_i^2.
\end{eqnarray}
Substituting $\Phi(z)=z^3$, the equation can be expressed as
\begin{eqnarray}
    \Phi(z_{0,i}+\delta z_i) = z_{0,i}^3 +3z_{0,i}^2\delta z_i+ 3z_{0,i}\delta z_i^2.
\end{eqnarray}
We can further expand $\delta z$ in terms of normal modes $\delta z_{\rm{c}}$ and $\delta z_{\rm{s}}$, denoting the displacement of the COM and stretch modes. We define the transformation into the normal-mode coordinates as
\begin{align}
    \delta z_{\rm{c}} &= \frac{1}{2}(\delta z_1 + \delta z_2) \nonumber \\
    \delta z_{\rm{s}} &= \delta z_1 - \delta z_2.
\end{align}
The second order term of the potential then becomes $3z_{0,i}(\delta z_{\rm{c}} \pm \delta z_{\rm{s}}/2)^2$ for electrons 1 and 2, respectively. After applying the time dependence $\cos(\omega_{\rm{p}}t)$, only the terms proportional to $\delta z_{\rm{c}}\delta z_{\rm{s}}$ remain, and all other terms can be dropped under RWA \cite{Hou2023IndirectCooling}. Therefore, the expression for $\Phi(z_{0,i}+\delta z_i)$ can be further simplified into
\begin{eqnarray}
    \Phi(z_{0,i}+\delta z_i) \approx 3z_{0,i}\alpha \delta z_{\rm{c}}\delta z_{\rm{s}},
\end{eqnarray}
where $\alpha=1,-1$ for $i=1,2$. 

The total potential energy of the two electrons within the potential $\Phi (\mbf{z}_0+\delta \mbf{z})$ is then the sum of the energies of both electrons, expressed as
\begin{eqnarray}
    U_{\rm{p}}(\mbf{z}_0+\delta \mbf{z}) = 3q\delta z_{\rm{c}}\delta z_{\rm{s}}(z_{0,1}-z_{0,2}).
\end{eqnarray}
Assuming the trapping potential in the axial direction is harmonic, we expect the equilibrium positions of the two electrons to be symmetric across $z=0$, hence $z_{0,1} = -z_{0,2}$, and the total energy can be rewritten as
\begin{eqnarray}\label{eq:E_p}
    U_{\rm{p}}(\mbf{z}_0+\delta \mbf{z}) = 3ql\delta z_{\rm{c}}\delta z_{\rm{s}},
\end{eqnarray}
where $l$ is the equilibrium distance between the two electrons. 

Substituting Eq.~\ref{eq:E_p} into Eq.~\ref{eq:U_z} and expressing the displacement operators in terms of annihilation and creation operators, after RWA the coupling Hamiltonian can be written as
 
\begin{eqnarray}
    H_I \approx \hbar \left(\frac{3lqV_2}{4mr_0^3\sqrt{\omega_{\rm{s}}\omega_{\rm{c}}}}\right)(\a{c}\adag{s}+\adag{c}\a{s}).
\end{eqnarray}
Therefore, we expect the COM mode to equilibrate to the same number of phonons as the stretch mode. If the COM mode can be cooled down to the tank circuit temperature at 0.4\,K, the stretch mode will be cooled down to $T_s = T_c(\omega_{\rm{s}}/\omega_{\rm{c}}) = 0.69$\,K. Notice that this final temperature is well below the threshold temperature for stable Wigner crystal formation with crystal lifetime of 1\,ms.

\begin{figure}
    \includegraphics[width=0.96\linewidth, left]{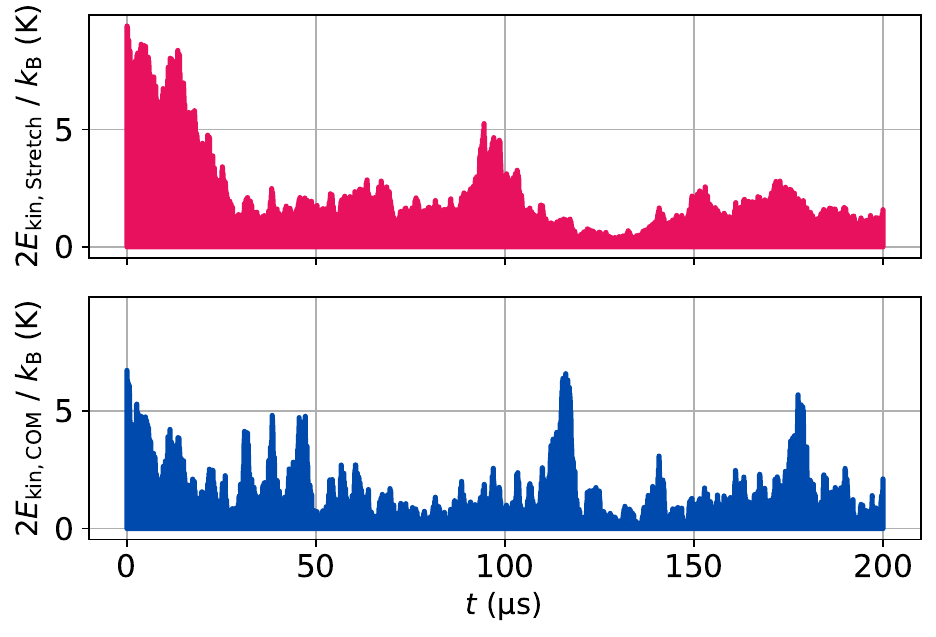}
    \caption{Simulation of cooling the two-electron stretch mode parametrically, with energy evolution shown for the axial stretch mode (top) and the axial COM mode (bottom). The COM mode is cooled through the coupling to the tank circuit, and the stretch mode is coupled to the COM mode through an external parametric RF potential.}
    \label{fig:cooling-two_electron}
\end{figure}

Here, we chose $V_2/r_0^3 = A_{\rm{p}}V_0/r_0^2$, where $V_0$ is the voltage applied for the RF trapping potential and $A_{\rm{p}}=10$ is selected to demonstrate the cooling effect within a reasonable simulation time. We conducted a simulation with the radial modes initialized at zero energies and the axial modes at 4K (COM) and 8K (stretch), below the crystallization thresholds. The resulting energy evolution of the axial COM and stretch mode is shown in Fig.~\ref{fig:cooling-two_electron}, where the equilibrium temperature of the COM and stretch modes are found to be 0.36\,K and 0.60\,K. The averaged values over 20 runs are found to be $0.41\pm 0.07$\,K and $0.81\pm 0.12$\,K, respectively for COM and stretch modes. The slight disagreement with the theoretical equilibrium temperatures can again be attributed to the randomness in the noise generation.
In addition, similar to the single-electron parametric cooling case, the primary mode to cool is the mode with lower frequency, hence the noise may cause a larger energy increase in the stretch mode when parametrically coupled. 

It is worth noting that it might not be possible to use this process to cool the stretch mode down from a highly energetic state. This can be seen from the potential expansion in Eq.~\ref{eq:cooling-expansion}, where the displacements of the two electrons are assumed to be relatively small compared to the equilibrium positions. This assumption does not hold when the electrons are in a cloud state. As a result, it is still necessary to first cool the individual electrons down to a low temperature and transport them with low heating.

\section{Influence of magnetic field} \label{sec:B field}
To utilize electrons in linear Paul traps for quantum information processing, we need to control the electron spins as qubits. To split the spin degeneracies of the electrons, we proposed to apply a static magnetic field of 3.6\,mT in the $y$ direction of the trap, with a corresponding electron cyclotron frequency of $\omega_{\rm{ce}}= eB/m=2\pi\times 100$\,MHz \cite{YuFeasibility}. 


The dynamics of charged particles in a linear Paul trap under an applied magnetic field have been studied in the context of combined Paul traps used for mass spectrometry \cite{li1992combinedPaul, huang1997combinedPaul}. In these studies, a static magnetic field is typically applied along the axial direction, thus producing additional confinement in the radial direction on top of the RF confinement in conventional Paul traps and reducing the impact of background collisions on the ion motion \cite{huang1997combinedPaul}. The applied magnetic field results in coupled equations of motion in the degenerate radial $x$ and $y$ directions. To simplify this, a time-dependent coordinate transformation was performed, moving to the rotating frame that effectively decouples radial motions. By comparing the resulting equations to the general form of the Mathieu equation, we can find that the applied magnetic field modifies the 
stability parameter $a_{x,y}$ to $a_{x,y}=(\omega_{\rm{ce}}^2-2\omega_z^2)/\wrf^2$ \cite{li1992combinedPaul}. Therefore, a large magnetic field changes $a_{x,y}$, influencing the stability of electrons in Paul traps. This effect is more prominent for electrons than for ions as the small mass of electrons increases the cyclotron frequency quickly to the point where it becomes comparable to the secular frequencies of the Paul trap. 

Since we plan to apply the magnetic field in the $y$ direction, the cyclotron motion will then be in the $x$-$z$ plane, where the oscillation modes do not share the same frequency. Because of this asymmetry, a coordinate transformation no longer simplifies the equations, necessitating numerical simulations to map out the accurate stability regions. 

\begin{figure*}
    \centering
    \includegraphics[width=0.96\linewidth]{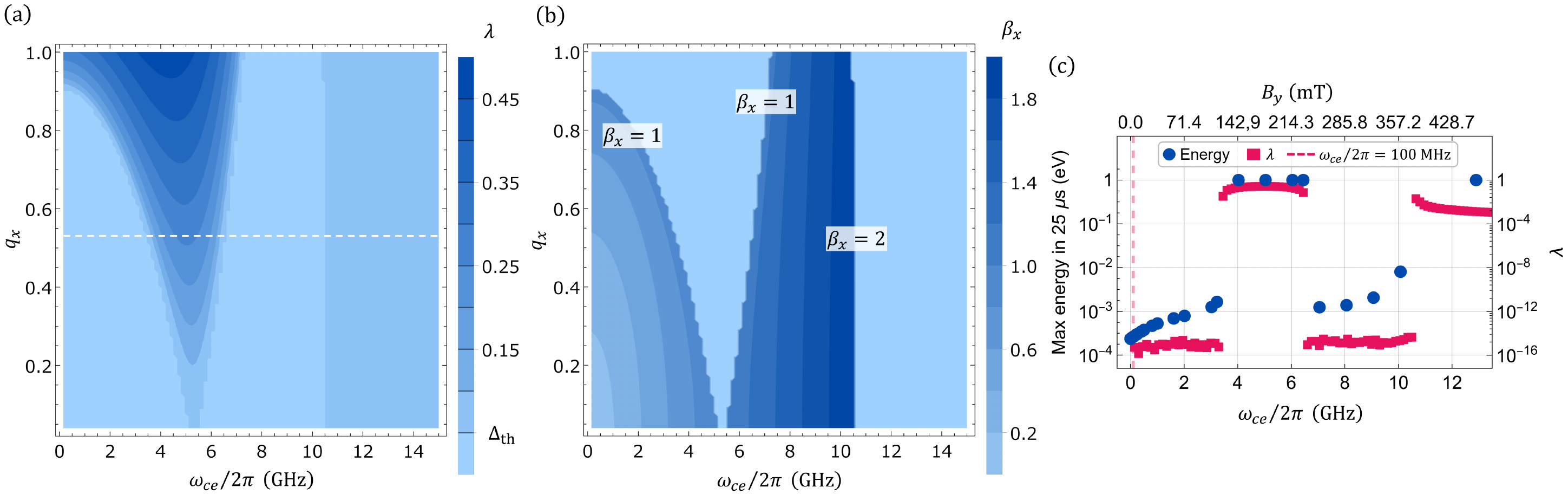}
    \caption{Stability diagram of electron motions under varying magnetic field strengths and trap frequencies. (a) The Floquet exponent $\lambda$ plotted against the Mathieu $q_x$ parameter, and the cyclotron frequency $\omega_{ce}/2 \pi$. Electron motion is unstable when $\lambda > \Delta_{\rm{th}}$. The white dashed line represents the proposed configuration with $\omega_r/2\pi=2$\,GHz and $q_x = 0.53$. (b) The value of $\beta_x = 2\omega_{x,\rm{tot}}/\wrf$ across stable regions for different $q_x$ and $\omega_{\rm{ce}}/2\pi$. The drive frequency is kept constant at $\wrf/2\pi = 10.6$\,GHz. The instability boundary in (a) matches the lines where $\beta_x$ takes integer values in (b), indicating that nonlinear resonances between the RF drive and the total secular frequency contribute to trap instability. (c) The comparison of the maximum radial total energy in a $25\,\text{\textmu s}$ trajectory of a single electron with $q_x=0.53$ and the $\lambda$ values against differently applied magnetic field strengths and corresponding cyclotron frequencies. The energies were extracted through the trajectories simulated with the RK3 algorithm, while the $\lambda$ values were taken from the line cut in (a). In the RK3 simulations, the electrons were initialized with kinetic energies equivalent to 0.4\,K in all modes, and the magnetic field was applied at the beginning of the simulation. The red dashed line represents the desired 100\,MHz frequency we proposed, corresponding to a magnetic field strength of 3.6\,mT. The maximum energy shown is capped at 1\,eV, corresponding to the typical trap depth.}
    \label{fig:res-B_field_HB}
\end{figure*}

We investigated the stability of a trapped electron for different trap frequencies by varying the $q_x$ parameter. Since we focused on the transition between a stable and unstable motion, the AFT-HB method was chosen to efficiently explore a wide range of parameters \cite{Junge2021, Nicks2024}. Using Floquet theory, we calculated the Floquet exponent $\lambda$ across different $q_x$ and $\omega_{\rm{ce}}$ values in Fig.~\ref{fig:res-B_field_HB}(a). In theory, the electron's motion is considered unstable when $\lambda > 0$, indicating the exponential growth in amplitude \cite{Detroux2015}. However, computing $\lambda$ always accompanies errors leading to a positive $\lambda$ even for the stable electron motion. Therefore, we define a threshold value $\Delta_{\mathrm{th}} = 10^{-10}$ resulting from the numerical accuracy limits and consider that the motion of the electron is stable when $\lambda < \Delta_{\mathrm{th}}$ is satisfied.

Our results show that stability depends on both the cyclotron frequency $\omega_{\rm{ce}}$ and $q_x$, consistent with previous findings in combined Paul trap studies \cite{huang1997combinedPaul}. To identify the cause of instability, we computed the total secular frequencies, incorporating both the trapping and magnetic field effects, and plotted the parameter $\beta_x = 2\omega_{x,tot}/\Omega_{\rm{rf}}$ in Fig.~\ref{fig:res-B_field_HB}(b), where $\omega_{x,tot}$ is the total secular frequency in $x$ direction. We found that the boundary regions where $\lambda>0$ closely match the lines where $\beta_x$ is an integer. This indicates that instabilities are caused by parametric resonances, which occur when the RF drive frequency is an integer multiple of the secular frequency.

To confirm the validity of the AFT-HB method, we compared the results with a time-domain trajectory simulation of a single electron using a fixed set of trap frequencies and varied the magnetic field strength. The energy evolution of the electron is monitored over a 25\,\textmu s simulation time, and the maximum energy during this period is recorded. If the maximum radial energy at any time of the trajectory exceeds 1\,eV, we consider the electron trapping to be unstable, since 1\,eV is comparable to a typical trap depth of our designs. As shown in Fig.~\ref{fig:res-B_field_HB}(c), we observed unstable trapping for cyclotron frequencies between 4\,GHz and 7\,GHz, and we also observed a second region of stability for cyclotron frequencies between 7\,GHz and 10\,GHz. The corresponding $\lambda$ value is also shown for the same frequency configuration as the time-domain simulations, and the stability region agrees with the time-domain results.

For our proposal, we are interested in the region of cyclotron frequencies below 1\,GHz. We found that the electron trapping is stable in this region. We also observed that the maximum energy in this stable region can still be greater than the energy at zero magnetic field. However, the energy does not increase over time and hence the electrons can still be cooled resistively. Some sample trajectories of the electron in stable and unstable regions are presented in Appendix \ref{sec:sample_traj_B_field}.


\section{Conclusion and outlook}
\label{sec:conclusion}
In this work, we set up a numerical framework to analyze the electron dynamics in linear Paul traps we designed, and the results obtained here provide essential information for future designs of the experimental apparatus. Particularly, we learned from the simulations that we need the design to be in a low-temperature cryogenic environment below 2\,K to allow the formation of Wigner crystals with lifetimes of 1\,ms. The applied static magnetic field needs to be below a certain threshold to ensure electron trapping stability. In addition, we also emulated real experimental processes by numerically demonstrating the feasibility of a proposed cooling scheme. This general methodology of Wigner crystal formation might not only be important for quantum information processing, but also for condensed matter or plasma physics where the phase transition and spin interactions are of interest.

This work highlights the near-term importance of designing an efficient strategy to cool the motional modes of electrons. Particularly, methods that can efficiently cool the electrons to their motional ground state \cite{Osada2022FeasibilitySystems} are beneficial to the platform, since the heating induced in the transportation process would be minimized and the formation of Wigner crystal would be easier. In addition, novel methods that can cool the motional modes to a temperature below the physical temperature of the cryogenic environment can be meaningful, since they can potentially relax the needs for an ultra-low temperature environment.

In the long run, it is also important to consider potential improvements on the design of our apparatus for quantum information processing. Since Wigner crystals can be formed more easily at higher trap frequencies, it can be beneficial to further increase the drive frequencies beyond 10\,GHz, which then requires a larger voltage amplitude and can cause many engineering challenges involving thermal loads, power consumptions, and device parasitic capacitance. These challenges may be mitigated by more advanced trap designs and fabrication techniques \cite{xu20233dTrap}, or using mm-wave electronics \cite{anferov2024mmWaveCQED} and superconducting electronics \cite{daniilidis2013ElectronSuperconductingTrap, tsuchimoto2024SuperconductingIOnTrap}. In addition, other two-qubit gate schemes that do not rely on the formation of Wigner crystals may be of interest, such as gates utilizing a double-well potential \cite{foot2011doubleWellGates, HarlanderInsbruck2011} or cavity-mediated coupling schemes \cite{mielke2023CavityGateQDots, welte2018CavityGateAtoms}.

\appendix
\section{Trajectories of crystal and cloud states} \label{sec:sample_traj_crystal}
\begin{figure}[ht]
    \includegraphics[width=0.93\linewidth, left]{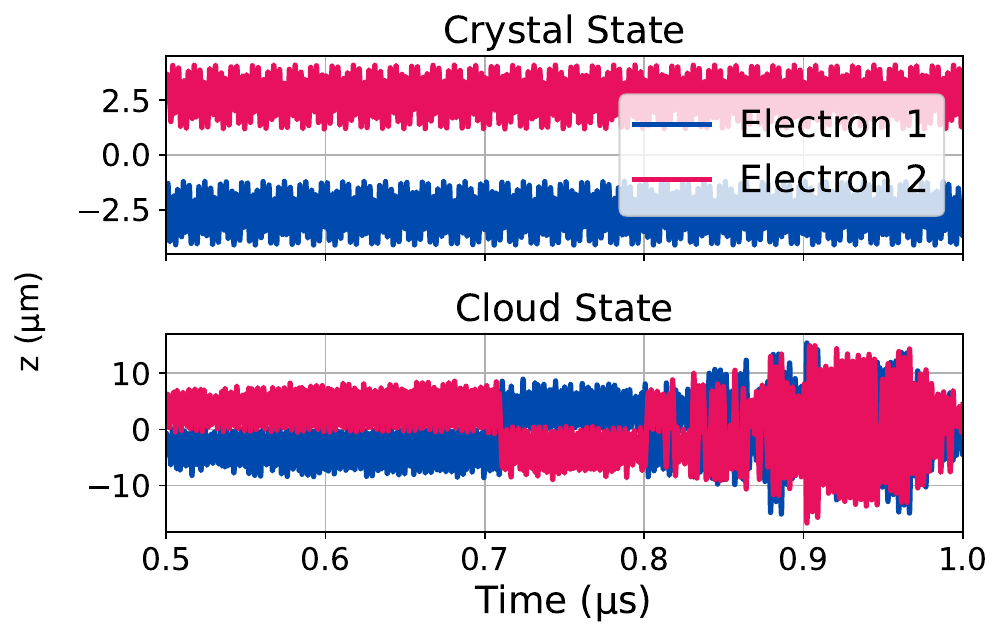}
    \caption{Sample trajectories along the axial directions for an electron crystal (top) and electron cloud (bottom) with axial stretch mode temperatures at $0.4$\,K and $18$\,K respectively. The other motional modes are at $0.4$\,K for both scenarios.}
    \label{fig:sup-sample_traj_crystal}
\end{figure}

Sample axial trajectories of two electrons in crystal and cloud states for a 1\,\textmu s time period are shown in Fig.~\ref{fig:sup-sample_traj_crystal}. For the crystal example, all motional modes have an assigned energy of 0.4\,K . The two electrons' trajectories in the axial direction do not have any overlap throughout the whole simulation. 
In comparison, we show here a second simulation of the cloud state with 18\,K axial stretch mode energy, and in this case the electrons are switching axial positions after a certain amount of time.

\section{Trajectories of single electrons under different applied magnetic fields} \label{sec:sample_traj_B_field}

\begin{figure}
    \centering
    \includegraphics[width=0.96\linewidth]{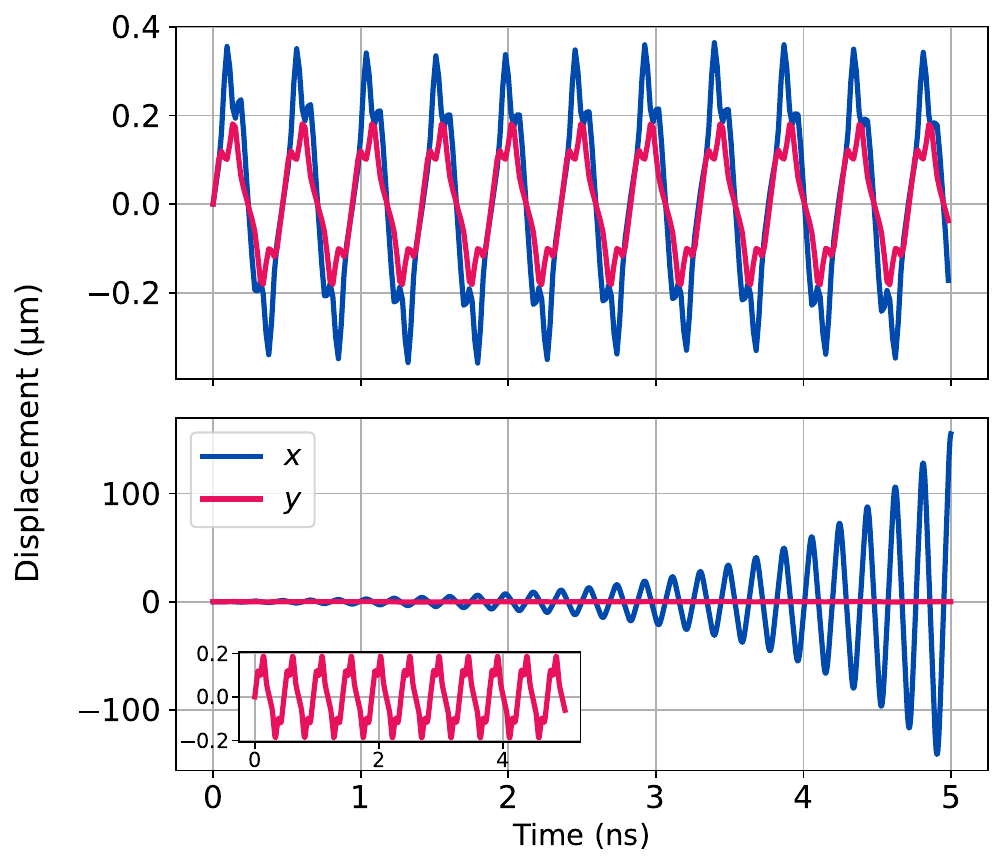}
    \caption{Sample trajectories for single electrons in $x$ (blue) and $y$ directions (red). (a) A 3.6\,mT magnetic field is applied, corresponding to a 100\,MHz cyclotron frequency, and the electron motion appears stable. The inset figure shows the trajectory for the first 5\,ns to show the details of the dynamics. (b) The unstable region with 125\,mT applied magnetic field, corresponding to a 3.5\,GHz cyclotron frequency. The motion in $x$ direction grows rapidly in amplitude and the motion in $y$ direction remains stable. The inset figure zooms into the trajectory along the $y$ direction to better show its dynamics.}
    \label{fig:sup-B_field_sample_trajectory}
\end{figure}

Sample trajectories for a single electron under different static magnetic field strengths are shown in Fig.~\ref{fig:sup-B_field_sample_trajectory}. Here, the displacements in the radial directions are shown to provide a good visualization, especially of the difference in dynamics in the $x$ and $y$ directions. We expect that when instability occurs, the displacement in $x$ and $z$ directions will increase rapidly due to the nonlinear resonances described in Sec.~\ref{sec:B field}, while the motion in $y$ direction will not be affected. This is because the static magnetic field is applied along the $y$ direction and there is no coupling between any directions, assuming a harmonic potential. Due to the exponential growth in the displacement, the electron's energy will quickly exceed 1\,eV, beyond which point we consider the electron is lost. 

\acknowledgments
A.H. and E.H. contributed equally to this work. We acknowledge the funding support from AFOSR. A.H. thanks Wenhan Hua for conversations regarding the modeling of Johnson noise, Dietrich Leibfried for insights on cooling and transporting electrons, and Jonathan Wurtele and Joel Fajans for intuitions related to the instability under magnetic fields. K.T. acknowledges JST SPRING (Grant No.JPMJSP2108) and A.N. acknowledges JST PRESTO (Grant number JPMJPR2258).

\bibliography{references}

\end{document}